# Porous silica beads produced by nanofluid emulsion freezing

Mathieu Nespoulous, Mickaël Antoni, Carine Chassigneux and Renaud Denoyel

*Aix-Marseille Université, CNRS, MADIREL UMR 7246, Marseille, France*

It is shown that porous spherical particles can be obtained via the freezing of silica nanoparticle aqueous suspensions emulsified in a continuous oil phase. After two freeze-thaw cycles, nanoparticles turn aggregated into flocculated objects with microstructure that depends upon emulsion volume fraction and droplet size. For low volume fractions, regular microspheres are produced while for large ones, irregular beads with several tens of micrometer radius are formed. Electronic microscopy, mercury porosimetry and nitrogen adsorption are used to get insights into these porous particles typical radius, pore size distribution, surface area and pore network structure. All exhibit mesopores that result from inter-nanoparticle spacing after flocculation. An unexpected macroporous domain appears which is not observed when drying non-emulsified suspensions. This macroporosity is interpreted as the signature of dendrite formation during the undercooled period, right before freezing occurs. Beside this additional macroporosity, the protocols presented in this article constitute also promising emulsion-based routes for porous material synthesis with original geometry, chemical composition and porosity.

**Corresponding author**: Mathieu Nespoulous (mathieu.nespoulous@univ-amu.fr)

# INTRODUCTION

Porous materials are ubiquitous with many applications in separation, purification or catalysis. Intense research activity has been devoted to their characterization for decades [1-4]. One still challenging problem is the synthesis of porous objects with controlled size and porosity. The production of spherically shaped microparticles is for example of first importance for separative analytical techniques, like chromatography, where transport properties of columns must be carefully controlled. In situ chemical templating techniques have been used to this end with silica non-porous spheres produced by the Stöber method [5]. Spherical porous materials are also commonly synthetized using polymers as templates. In addition to molecules, latex particles have been used as template to produce small mesoporous spheres with radius in the range 200-800 nm while the pore size is controlled by triblock copolymers [6]. In industrial processes, spherical porous particles have long been produced by spray drying methods with suspensions of non-porous particles [7]. Several levels of porosity can be achieved depending upon the properties of the initial suspensions. The use of mesoporous particle suspensions allows for example the preparation of porous spheres with two typical porosity scales [8].

Besides spray drying, emulsion-based synthesis of porous materials with particles has also drawn a considerable attention over past years [9,10]. Emulsified suspensions are known to involve complex adsorption and wetting mechanisms. It has been for example demonstrated already more than a century ago by Ramsden and Pickering that such emulsions can be stabilized only by surface active particles [11,12]. In this context, intense activity has been devoted to nanofluid emulsions that consist of emulsions where particles are nanometer sized and dispersed in either the continuous or the dispersed phase. Surfactant molecules have been further used to tune interfacial properties of these nanoparticles. Studies on synergistic mechanisms [13], liquid/liquid interfacial properties of drops [14] as well as emulsion rheology [15] and drop morphology [16-18] have been reported for a large variety of nanoparticles and surfactant molecules. In the case of silica nanoparticles, both oil in water and water in oil nanofluid emulsions have been studied for porous material elaboration. For oil in water emulsions, it has been evidenced that such nanoparticles get polymerized along the drop interfaces [9] while for water in oil ones, mesoporous silica spheres were obtained [19]. Hierarchical porosity systems can be further elaborated by multiple O/W/O emulsions where the dispersed oil droplets create macropores surrounded by mesopores produced by silica polymerization [20].

In the context of the present work, nanofluid emulsions have the great interest to offer a broad range of possibilities for sol-gel preparation of porous materials in well controlled conditions. But one drawback here is the use of large additives concentrations, mainly surfactant molecules, for chemical templating that have to be eliminated by calcination and/or complex washing. Time consuming procedures, not always easy to manage regarding environmental standards, are therefore necessary. Recent studies have shown that this problem can be overcome using foaming techniques or ice templating [21]. This latter has been successfully used to produce controlled porous polymeric [22, 23] or inorganic [24] materials with very limited amounts of additives and a rather reduced use of solvents. Ice templating has been used for example to structure porous polymers in the case of water in oil emulsions, where both phases contain polymers [25]. The ice templating may also occur in both water and oil phases: the porous spheres, produced in oil droplets, are then obtained after dissolution of the porous polymer produced in the water continuous phase [26]. The synthesis of macro-mesoporous inorganic powders with the principle of ice templating has also been addressed in the case of ceramic microspheres [27] but using cyclohexane as freezing solvent and high temperature subsequent treatments. To our knowledge, the use of ice templating, to synthesize porous inorganic powders starting directly from a suspension of nanoparticles in oil emulsion, has not been developed.

The aim of this work is to contribute to fill this gap. The formation of porous structures has been recently observed when applying freeze-thaw cycles to nanoparticle suspensions in a pendent drop configuration [28]. Based on this observation, it is proposed hereafter to use aqueous nanofluids of silica nanoparticles emulsified in a heptane continuous phase with SPAN80 as stabilizer. Micrometer sized droplets are easily formed in such systems. Different emulsion volume fractions are studied for the same emulsification protocol and yield modifications in both droplet sizes and shapes. After drying, spherical micrometric particles consisting of flocculated silica nanoparticles are evidenced for low volume fraction emulsions. For large volume fractions, flocculated particles turn deformed and exhibit unexpected macroporosity. Both cases are discussed at the light of known water freezing mechanisms. Conversely to conventional sol-gel chemical synthesis, the production of flocculated particles in this work can be seen as the result of a physical synthesis. Minimum amount of chemicals is necessary and almost all the solvent can be reused for subsequent synthesis rounds.

## EXPERIMENTAL

**Emulsion preparation** – The continuous phase of the emulsions is a solution of heptane (Sigma-Aldrich) with SPAN80 (Sigma-Aldrich) at 10 g/L concentration. The dispersed phase is an aqueous nanofluid of amorphous silica nanoparticles (Ludox SM 30 wt.% or 16% volume, provided by Sigma-Aldrich). The latter was used without further purification and dilution. 10 mL samples are prepared in 16 mL glass vials by introducing first the heptane/SPAN80 solution and then, on the top of it, the nanofluid with a micropipette (see

Figure 1(a)). Five different emulsion volume fractions, $\phi$ = 1%, 2%, 5%, 10% and 20%, are prepared: this is the ratio of nanofluid introduced volume to the total volume. W/O emulsions are finally created with a high-speed homogenizer (IKA T18 Ultra-Turrax) for 2 minutes at 8000 rpm (see Figure 1(b)). After mixing, the emulsion is immediately introduced in a freezing chamber.

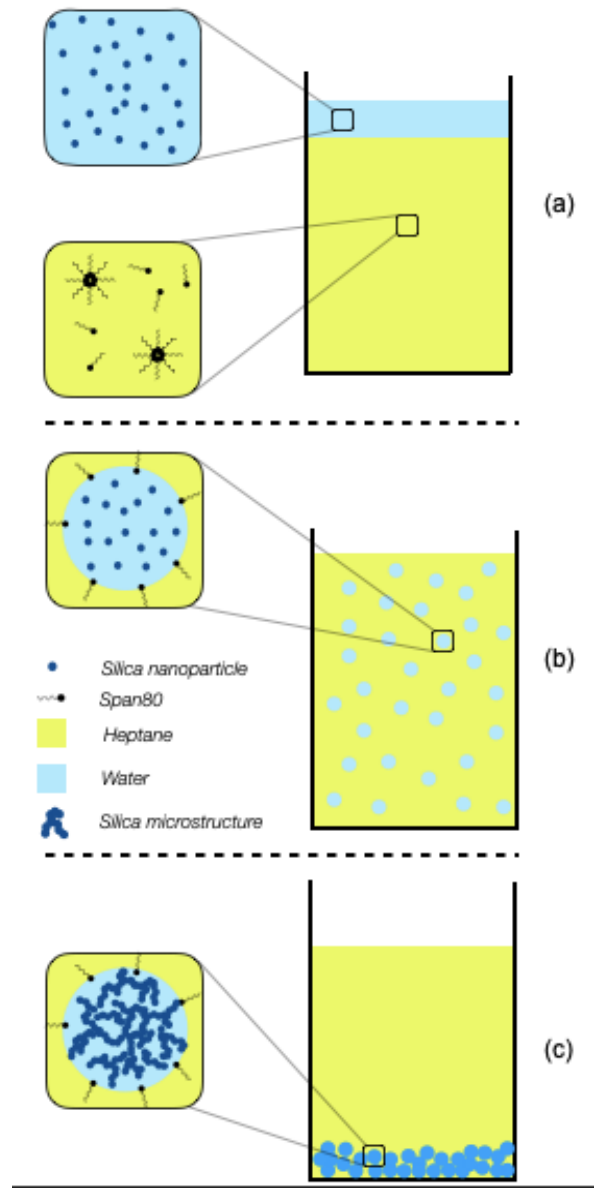

Figure 1: Preparation of porous silica microparticles by freeze-thawing a W/O emulsion. (a) heptane/SPAN80 and nanofluid solution preparation. (b) emulsion creation by Ultra-Turrax homogenization. (c) sedimentation of flocculated silica microparticles after freezing.

**Freezing and drying** – Two freeze-thaw cycles are applied with a SP Scientific VirTis Sentry 2.0 lyophilizer freezing chamber. The temperature is cycled at an average cooling rate of 2 °C/min from room temperature down to -40 °C and back to room temperature. At -40 °C the dispersed aqueous nanofluid is frozen whereas the heptane continuous phase remains liquid (as the heptane freezing is about -91 °C). After the two freeze-thaw cycles, flocculated particles, denoted as beads in the following, are created and sediment in a bed at the bottom of the vials (see Figure 1(c)). Most of the heptane is then carefully removed with a Pasteur pipette before sample drying in an oven at 80 °C overnight.

**Mercury porosimetry** – Pore size distribution is determined with a Quantachrome Poremaster apparatus. Applied pressure runs from 4 kPa to 348 MPa and a contact angle of $\theta$ = 140° is used as recommended for silica-mercury interface [29]. This allows to probe pore diameters ranging from 300 µm down to 4 nm. Two pressure domains are successively explored: low pressure (4 to 345 kPa) and high pressure (138 kPa to 345 MPa). Pore diameter is computed from the intrusion experiments with the Laplace-Washburn equation. The mass of sample is about 100 mg and experiment temperature 22 °C.

**Nitrogen sorption** – The surface area and mesopore size distribution of the beads are derived from gas adsorption analysis, using a Micromeritics ASAP 2010 apparatus. The samples, again of mass close to 100 mg, were first outgassed in stable vacuum (~0.05 mbar) conditions at 150 °C for 12h. $N_2$ adsorption/desorption isotherms are then measured at 77 K.

**Scanning electron microscopy** (SEM) – Microstructure morphology is studied using a Philips XL30 SFEG STEM scanning electron microscope associated to an Oxford energy-dispersive spectrometry analysis system. Samples are carefully deposited on a carbon film before gold-palladium metallization.

## RESULTS AND DISCUSSION

Because it is known that the size of droplets has an influence on the freezing mechanism [35] and that emulsion volume fraction has an influence on droplet size, the influence of emulsion volume fraction is the main investigated parameter of this study. Figure 2 presents typical SEM images of samples obtained from the original suspension (*i.e.* without emulsification) and from five emulsified samples differing only by their emulsion volume fraction $\phi$. In the absence of emulsification, irregular aggregates with sharp edges are produced (top left image of Fig. 2) while for emulsified ones polydispersed beads with typical size increasing with volume fraction are evidenced. They consist of silica nanoparticles that have been aggregated by ice templating. For $\phi$ = 1% and 2% they appear as micrometer sized spheres with radius between 1 µm and 20 µm. For larger volume fractions, they look like deformed spheres with radius up to few tens of micrometer. Detailed visual inspection of the images for $\phi$ > 2% indicates that micrometer sized beads are still present and exhibit again a regular spherical geometry. Such beads are observed in all experiments and constitute the signature of the presence of sufficiently small droplets in the original emulsion for capillary forces to prevail and therefore preserve sphericity. Figure 2 also emphasizes that beads seem to organize into clusters. This is most likely a consequence of the drying procedure during which uncontrolled capillary forces may come into play as wetting films vanish in the very last evaporation phase. The way beads are deposited on the carbon film before SEM analysis has probably also to be accounted for here.

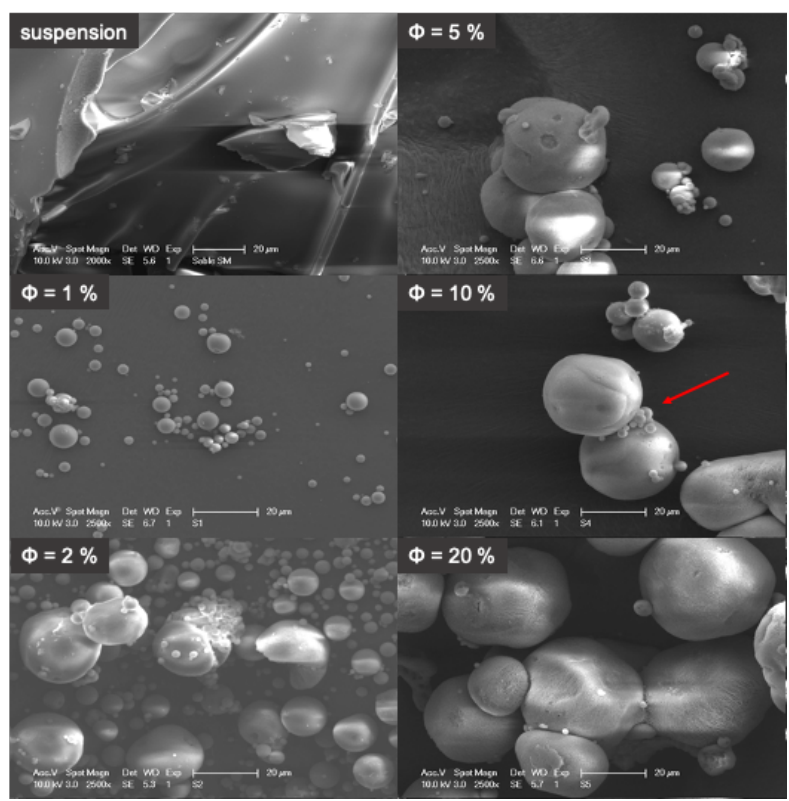

Figure 2: SEM micrographs of beads obtained after freezing as a function of $\phi$ (given top left in each image). The top left picture shows silica structures obtained from the freezing of a non-emulsified suspension. The red arrow in the picture for $\phi$ = 10% points to the region of interest emphasized in Figure 3.

All the experiments are performed with the original Ludox 30 wt.% dispersion. The reason for this is to make sure the densification processes involved in ice templating will create beads sufficiently stiff to keep memory of the droplets from which they were produced, as suggested by a preceding study with drops [28]. For diluted Ludox solutions, beads like the ones shown in Figure 2 would be instable and collapse under their own weight in the drying phase. Two freeze-thaw cycles are necessary here since only unconnected micrometer sized flocs are formed after the first cycle [28]. Optical microscopy images of the main features associated to each cycle is illustrated in figure 3 in a pendent drop configuration. Micrometer flocs are visible in figure 3(b). The flocs completely fill the droplets and show up as a micro-suspension of floating filamentous aggregates that sediment as time runs. Applying right after the first cycle a second one acts similarly but, instead of nanoparticles aggregation, micrometer sized flocs now aggregate and come compacted together. This procedure prevents sedimentation to set in and sufficient compactness for stable bead production.

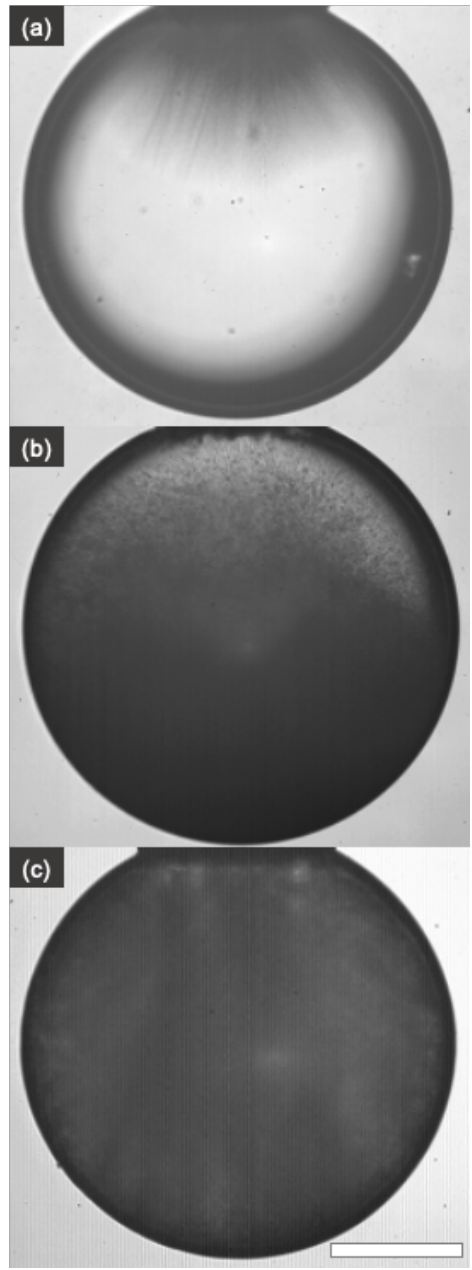

Figure 3: Optical microscopy images of a nanofluid pendent drop during the freeze-thawing procedure. (a) dendritic-freezing regime. (b) After first thawing – flocs are visible in the upper section of the drop. (c) Drop shape in its solid state after second thawing. Scale bar is 250 μm.

Figure 4 shows both small and large beads and illustrates the microstructures that will be characterized hereafter. One striking feature is the difference in the apparent surface textures. Micrometer sized beads turn compact and smooth while large ones exhibit a porous surface. The structure of the porous network therefore depends upon the size of droplets. Microstructures are known to be generated when solidification occurs in undercooled silica colloidal suspensions in pendent drop configurations [28]. For such undercooled systems, two successive freezing regimes have been evidenced: a dendritic growth regime that lasts for less than few milliseconds and subsequently a regime where latent heat is evacuated. This latter takes few seconds to be completed, in the case of millimetric droplets, and exhibits a spherical ice front propagating from outside to inside the drop. As just discussed, micrometric flocs are formed after the first freeze-thaw sequence and turn further flocculated with a second freeze-thaw sequence. Beads of the size of the pendent drop were produced this way and shown a porous surface as well, but the typical size was in the range of a few micrometers. Comparison of these results to the ones presented here for polydisperse emulsions clearly indicates that droplet size influences strongly the apparent porous structure of the material.

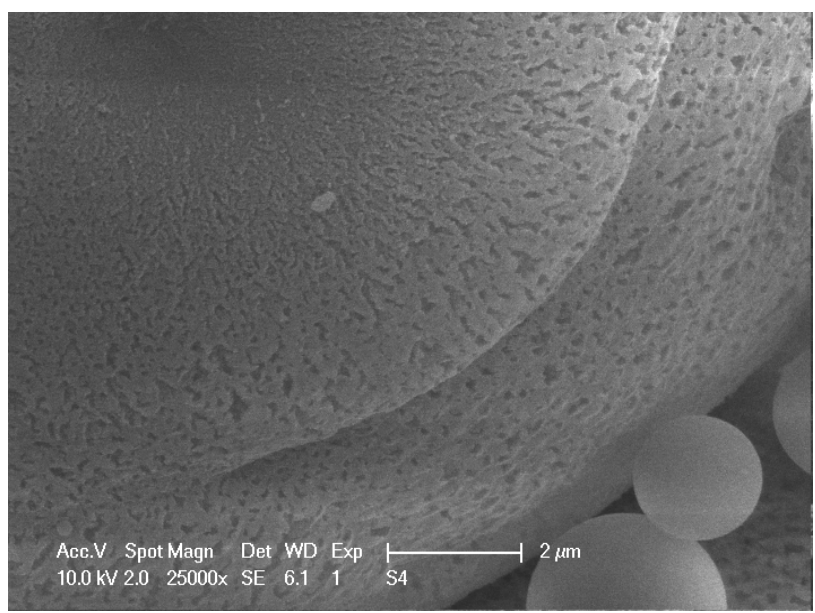

Figure 4: Zoom in the area emphasized by the arrow of Figure 2 ($\phi$ = 10 %). Small beads appear as compact objects while the large ones exhibit macroporosity.

The pore size distribution of the beads is evaluated by mercury porosimetry. Intrusion and extrusion curves for $\phi$ = 20 % are displayed in Figure 5. This figure also presents the curves of a reference Ludox sample obtained from a non-emulsified Ludox sample but that has undergone the same freeze-thaw cycles and drying procedures than the emulsified ones. At low pressures (*i.e.* large values of pore size *d*), a macroporosity shows up for both samples in the range *d* > 1 µm. This porosity is the usual signature of the compaction and the filling of interparticle space and corresponds here to the inter bead (IB) porosity. The corresponding porous volume is denoted as $V_{IB}$ hereafter. For the highest pressures (*i.e.* smallest values of *d*) mesopores are detected with typical diameter *d* ≈ 9 nm (resp. 5 nm) for the emulsion (resp. reference sample). This is again an interparticle signature but it now results from inter-nanoparticles (INP) residual space with porous volume given by $V_{INP}$. Surprisingly, an additional macroporosity in the range 50 < *d* < 200 nm is observable but only for emulsified samples. It is probably reminiscent of the size and shape of all the ice dendrites that propagated throughout the nanofluid as freezing took place. This specific bead porosity (B) has volume $V_B$ and was already visible on the surface of the large beads in Figure 4.

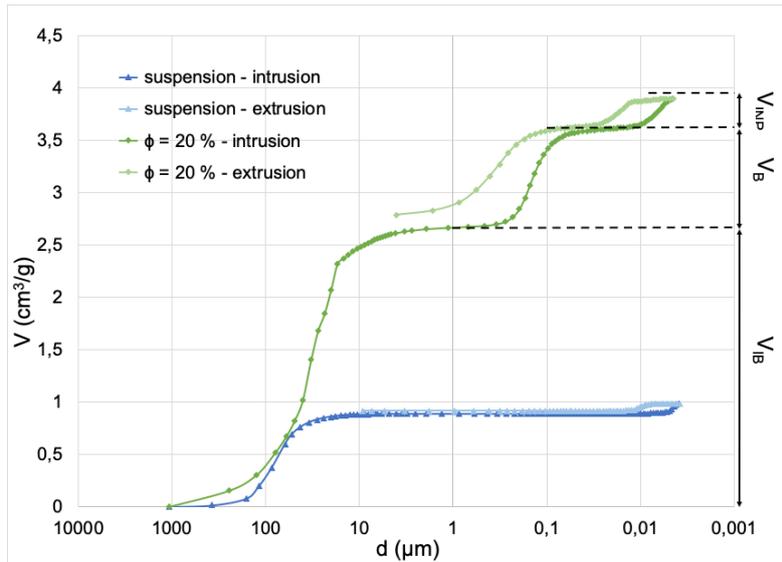

Figure 5: Cumulative intruded and extruded mercury volume as a function of pore diameter. Diamonds (resp. triangles) correspond to emulsion sample (resp. reference Ludox sample). Dashed lines qualitatively illustrate porous volumes $V_{IB}$, $V_B$ and $V_{INP}$ presented on the right of this figure.

Porous volumes $V_{IB}$, $V_B$ and $V_{INP}$ are each associated to a specific porosity that actually depends upon $\phi$. One question raised by the new macroporous volume $V_B$ evidenced in Figure 5 is the way it is modified when changing $\phi$. Mercury intrusion curves as a function of $\phi$ are plotted in Figure 6(a) and reveal the same three-step behavior than the one of Figure 5. They also reveal that $V_{IB}$ increases with $\phi$. From total intruded volume, silica density and $V_{IB}$, it is possible to calculate the apparent densities of beads that are shown to decrease with $\phi$, which is the reason for the increase of $V_{IB}$. The intrusion step in the IB range shifts moreover towards smaller values of $d$ when decreasing $\phi$ and indicates that bead typical size also decreases as already suggested by the SEM images of Figure 2. A careful image processing could obviously help to complement this result, but it is rather clear from this figure that different typical bead sizes are already observable when comparing the images for $\phi$ = 1% and 2% with the ones for $\phi$ = 10% and 20%.

Figure 6(b) shows intrusion-extrusion curves for $d$ < 1 µm and illustrates the two steps corresponding to B and INP porosity. Extrusion curves are interesting since they provide fundamental inputs about the topology and robustness of the porous network. The shape of the hysteresis changes with $\phi$. For $\phi$ = 10% and 20%, it displays a closure point. B and INP porosity domains are therefore well separated and reveal a two-level hierarchical structure with porous domains filled and emptied independently. For smaller values of $\phi$ however, the separation between the two domains is less marked: for example, macropores could be connected to outside through mesopores which act as constrictions. This trend demonstrates that the structure of the porous network is strongly modified and signs a change of connectivity of the network topology.

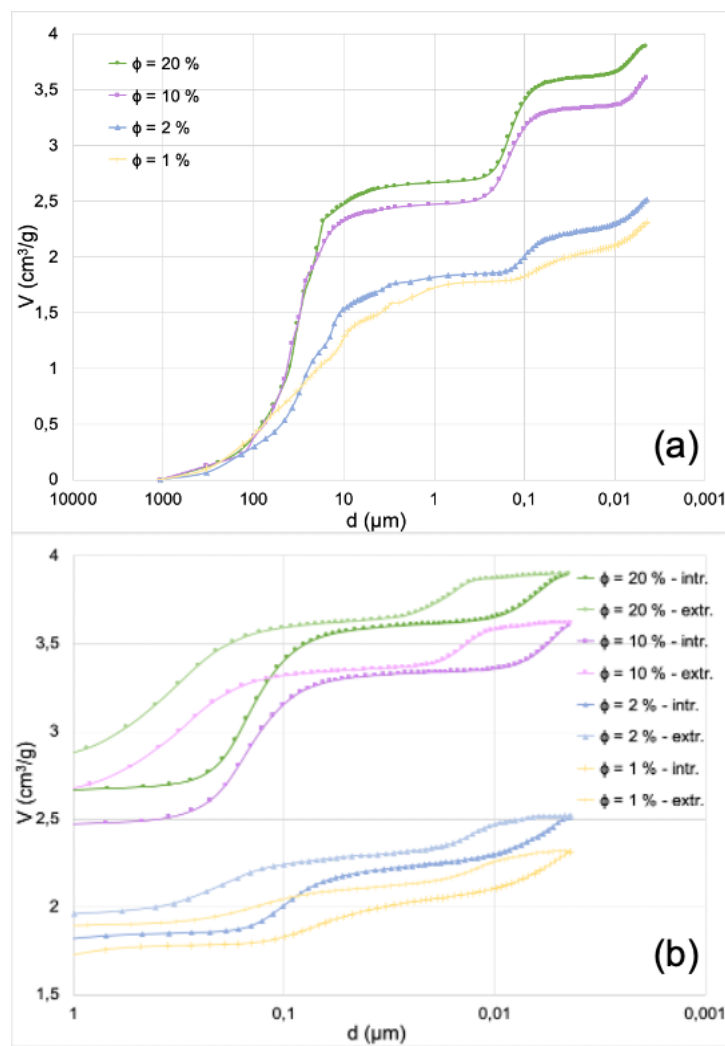

Figure 6: (a) Cumulative intruded mercury volume as a function of pore diameter and volume fraction $\phi$. (b) intrusion and extrusion curves for $d$ < 1 μm.

The inflexion points of the mercury cumulative intrusion curves are now further used to compute $V_B$ and the mean pore size $d_B$ in porous domain B as a function of $\phi$. This latter is given by the value of $d$ at the inflexion point within B. Intrusion curves are smoothed with a three-points running average prior to computations. Results are reported in figure 7. Both $V_B$ and $d_B$ exhibit a clear increasing trend with $d_B \approx 50$ nm for $\phi = 1\%$ and $d_B \approx 140$ nm for $\phi = 10\%$. Beyond this value, $d_B$ seems to have reached a plateau. As discussed above, the emulsification protocol allows the formation of larger droplets when $\phi$ is increased. The increasing value of $d_B$ with $\phi$ therefore suggests an increasing typical size of the dendrites created in the freezing phase. This result, together with the smaller fraction of constrictions in the pores geometry just discussed, confirm the key role played by droplet size in the final pore organization. As smaller droplets support more complex porous networks, other contributions like capillarity or the achievement of larger supercooling rates for nucleation, are most likely key in the overall ice templating mechanisms occurring in the emulsion droplets.

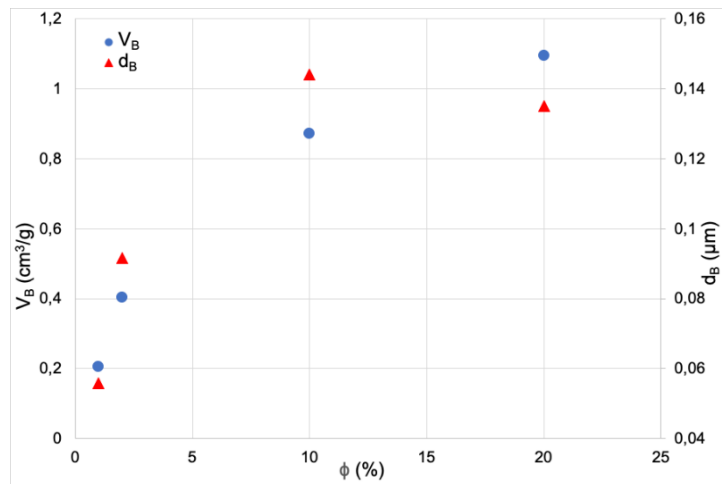

Figure 7: Bead porous volume $V_B$ and typical pore size $d_B$ in macroporous domain B as a function of $\phi$.

The non-saturated intrusion curves of Figure 6(a) for the smallest accessible pores (*i.e.* $d = 4{,}4$ nm) lead to a systematic underestimation of $V_{INP}$ (see Figure 5) and therefore prevents an accurate description of mesoporosity. To overcome this problem, further characterization is achieved with nitrogen adsorption at 77 K. Adsorption/desorption isotherms are displayed in Figure 8(a) for the reference sample and three emulsified ones with $\phi$ = 1%, $\phi$ = 10% and 20%. All exhibit type IV isotherms with typical hysteresis due to capillary condensation. The specific surface area is 227 m$^2$/g for $\phi$ = 1%, 10% and 20% and 225 m$^2$/g for the reference sample. The pore size distributions deduced from the BJH method [31] are shown in Figure 8(b).

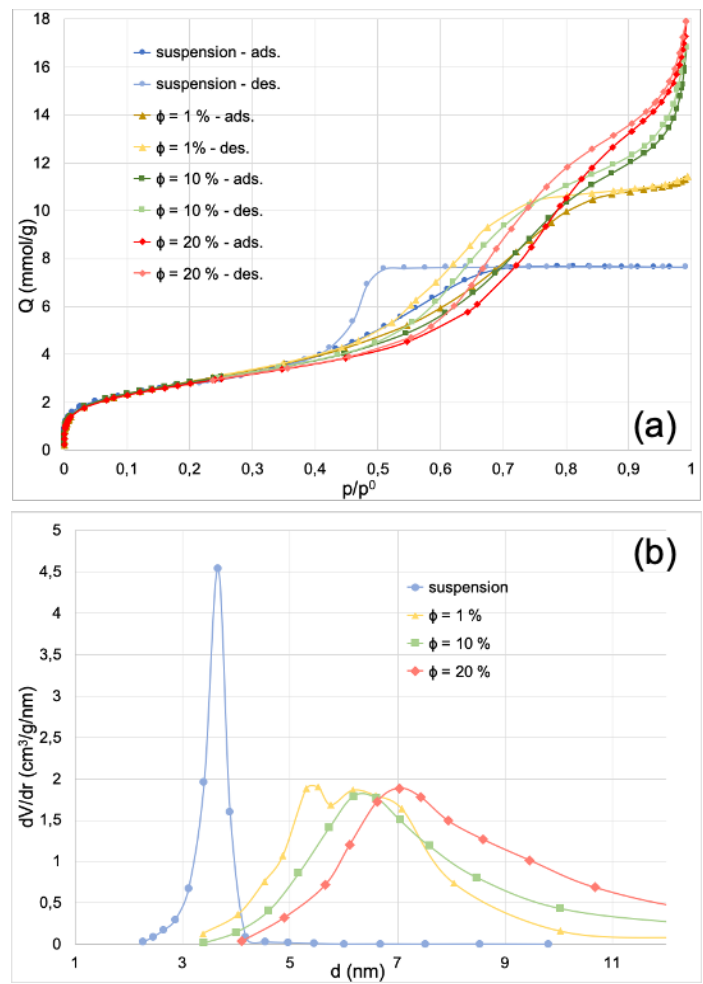

Figure 8: (a) Nitrogen sorption isotherms at 77 K for $\phi$ = 1%, $\phi$ = 10% and $\phi$ = 20% and for the reference Ludox sample. (b) Pore size distributions obtained via the BJH method.

For the reference Ludox sample, the shape of the hysteresis belongs to IV(a) type with H2(a) loop [32] and is characteristic of densely packed spheres [33]. The pore size obtained from BJH method applied to the desorption branch (see Figure 8(b)) yields a typical diameter for inter-nanoparticles pores of about $d_{INP} \approx 3.6$ nm that turns consistent with densely packed spheres of 15 nm diameter [34]. This result must however be handled carefully since the BJH method is not fully accurate close to the cavitation domain (corresponding to a relative pressure range $p/p°$ between 0.4 and 0.5 where $p°$ is the saturation pressure) and may under evaluate $d_{INP}$. The intraparticle porosity, calculated from the saturation plateau and the amorphous silica density (2.2 g/mL), is around 0.37 which is compatible with the 0.36 value of the random close packing in three dimensions [35]. In the case of emulsified samples, the hysteresis occurs at rather higher relative pressures showing larger pores: $d_{INP}$ = 6.1 nm for $\phi$ = 1%, $d_{INP}$ = 6.2 nm for $\phi$ = 10% and $d_{INP}$ = 7.0 nm for $\phi$ = 20%. This confirms that freezing generates flocculated structures with smaller densities, as already underlined with mercury porosimetry. Beads therefore present smaller number of close neighbors around each nanoparticle. Even though the saturation plateau at high relative pressures is poorly defined, an apparent porosity can still be calculated from the adsorbed nitrogen amount at the inflexion point that lies close to $p/p°$ = 0.94. One gets about 0.46 for $\phi$ = 1%, 0.49 for $\phi$ = 10% and 0.52 for $\phi$ = 20%.

The picture that shows up in this work for the beads structure is a first porous domain generated by dendrite growth surrounded by porous walls with both pore sizes depending upon droplet size. The important output of this study is the highlighting of macroporous domain B which does not exist when the nanofluid is not emulsified. This domain is a consequence of the freezing mechanisms taking place in the supercooled sequence of the nanofluid. It has been discussed above that millimeter sized pendent drops generate porous spheres having the same size than the original drop and with an apparent surface pore size of a few micrometers [28]. These latter are attributed to the "memory" of dendrites that are formed during the initial freezing step (see graphical abstract of [28]). This observation is to be compared with the surface pore sizes below 0.5 µm evidenced in the present study for the largest beads and to the fact that, for the smallest ones, pores are below the SEM resolution (see Figure 2). The freezing mechanisms in the case of pendent drops are moreover known to involve two successive steps: first a very fast formation of ice dendrites that invade the droplets followed by an advancing ice front from outside to inside the droplets. This latter is related to bulk solidification and heat transfer. In the case of small droplets of pure water, it has been further shown that the time delay between these two sequences is reduced. As heat exchanges with the surrounding continuous phase is eased for small droplets as the ones produced in emulsions, dendrite and bulk freezing sequences become most likely simultaneous. Such a transition from a two-step freezing to a single step one has been evidenced in the case of pure water droplets in air with a critical size between 0.1 and 1 µm [30]. The present work suggests the existence of a similar relationship between the size of the emulsion droplets and the size of the dendrites they can support. One important perspective of this study is to proceed with the understanding of this transition in the case of emulsified nanofluid suspensions.

CONCLUSION

Silica beads have been synthetized via the freezing and drying of nanofluid emulsions. Their shape, size and porous microstructure are shown to depend upon volume fraction. Porosity results from a two round ice templating. The porous structure of the beads consists of a macroporous domain and a mesoporous domain. This double-scale structure is the signature of the ice dendrite network that invaded the droplets in the first freezing sequence and of inter-nanoparticle spacing created in the subsequent flocculation phase. The two-level porosity increases its compactness and complexity as volume fraction is reduced: both macropore and mesopore size reduce while the connectivity is modified. The improvement of the control of droplet size in the emulsification procedure is one important perspective of this work in particular to produce materials with better controlled porosity, particle shape and particle size. A key point is to study the emulsification parameters such as stirring speed. The simple methodology used also opens new synthesis routes to incorporate active ingredients in porous structures, not by post modification, but by including them directly in the emulsified nanofluid before freezing. This work could therefore lead to potential innovative approaches for target applications like catalysis or drug delivery. Finally, the proposed material is the result of a physical synthesis involving a minimum amount of chemicals and almost all the solvent can be reused for subsequent synthesis rounds.

AKNOWLEDGMENTS

The authors would like to gratefully acknowledge ESA (ESA/MAP/EDDI), CNES (project Stabilité des dispersions - mesures et modélisation) for financial support and GdR/CNRS microgravité fondamentale et appliquée.